\documentstyle[aps,prd,preprint,epsfig,tighten]{revtex}
\begin{document}
\draft
\title{		Super-Kamiokande atmospheric neutrinos:\\ 
		Status of subdominant oscillations}
\author{ 	G.L.\ Fogli, E.\ Lisi, and A.\ Marrone}
\address{	Dipartimento di Fisica and Sezione INFN di Bari\\
             	Via Amendola 173, I-70126 Bari, Italy \\ }
\maketitle
\begin{abstract}
In the context of the recent (79.5 kTy) Super-Kamiokande atmospheric neutrino
data, we concisely review the status of  muonic-tauonic flavor oscillations and
of the subdominant electron or sterile neutrino mixing, in schemes with three
or four families and one dominant mass scale. In the three-family case, where
we include the full CHOOZ spectral data, we also show, through a specific
example, that ``maximal'' violations of the one-dominant mass scale
approximation are not ruled out yet.
\end{abstract}
\medskip
\pacs{PACS: 14.60.Pq, 13.15.+g, 95.85.Ry}

\section{Introduction}

The Super-Kamiokande (SK) Collaboration has recently presented an updated set
of atmospheric neutrino data for a detector exposure  of 79.5 kTy \cite{re01}.
The corresponding statistics is about twice as large as compared to the one
considered in our earlier published analysis of three-flavor mixing \cite{re02}
and of two-flavor mixing with nonstandard dynamics \cite{re03}, and is about
12\% larger than in our previous analysis of four-family schemes in
\cite{re04}. In addition, the CHOOZ collaboration presented in \cite{re05}
final  {\em spectral} data, which we now include in $3\nu$ analyses
\cite{re06}, improving the accuracy of our previous results \cite{re02} based
on the CHOOZ {\em total rate\/} \cite{re07}.

Therefore, we think it useful to present a concise update of our  $2\nu$,
$3\nu$, and $4\nu$ oscillation studies, performed under the hypothesis of one
mass-scale dominance, so as to elucidate the current status and implications of
subdominant (electron or sterile) neutrino mixing in $SK$. Finally, we also
discuss a specific $3\nu$ example which maximally violates the assumption of
one-dominant mass scale.

\section{Two neutrinos}

The standard $2\nu$ case of $\nu_\mu\leftrightarrow\nu_\tau$ oscillations can
be parametrized by one squared mass difference between two states
$(\nu_1,\nu_2)$,
\begin{equation}
m^2=|m^2_2-m^2_1|\ ,
\end{equation}
and by one mixing angle $\psi$, describing the flavor content of $\nu_2$,
\begin{eqnarray}
|\langle \nu_2|\nu_\mu\rangle|&=&s_\psi\ ,\\
|\langle \nu_2|\nu_\tau\rangle|&=&c_\psi\ ,
\end{eqnarray}
where $s=\sin$ and $c=\cos$.

Figure~1 shows our $2\nu$ best fit to the latest SK data, reached at
$(m^2,\sin^22\psi)=(3\times 10^{-3}{\rm\ eV}^2,0.97)$, and corresponding to
$\chi^2_{\min}=38.5$ for $55-2$ degrees of freedom---a very good fit.%
\footnote{The overall reduction of $\chi^2_{\min}$ with respect to the $2\nu$
(subcase) analysis presented in \protect\cite{re04} is mainly due to a better
agreement of the latest SK electron distributions with their no-oscillation
expectations (especially in the multi-GeV sample).}
The SK collaboration finds the $\chi^2$ minimum at a slightly different  point,
$(m^2,\sin^22\psi)_{\rm SK}=(2.5\times 10^{-3}{\rm\ eV}^2,1.0)$ \cite{re01}.
However, the  difference is not statistically significant, since the $\chi^2$
function turns out to be rather flat around the minimum, and the ``distance''
between our best-fit point and the SK one is only about one unit in $\Delta
\chi^2$. Concerning the $2\nu$ bounds on $(m^2,\psi)$ from parameter
estimation, they will be discussed later as limits of  $3\nu$ and $4\nu$ cases.

The striking evidence in favor of standard $\nu_\mu\leftrightarrow\nu_\tau$
oscillations (Fig.~1) strongly constrains nonstandard explanations. By using
the approach described in \cite{re03}, we parametrize a wide class of scenarios
involving nonstandard dynamics through three (free) parameters: an oscillation
amplitude $\alpha$, an overall phase factor $\beta$, and an energy exponent
$n$, the standard mass-mixing dynamics being recovered for $n=-1$.

Figure~2 shows the results of such a three-parameter fit in terms of the
projection $\chi^2(n)$. The corresponding bounds on $n$ give $n=-1.03\pm 0.31$
at 90\% C.L.\ ($\Delta \chi^2=6.25$ for three free parameters, $N_{\rm DF}=3$),
in perfect agreement with the standard case. Such results strengthen our
previous bounds obtained with smaller (45 kTy) SK statistics  ($n=0.9\pm 0.4$
\cite{re03}), and definitely exclude nonstandard dynamics with integer $n\neq
-1$ in the $\nu_\mu\leftrightarrow\nu_\tau$ channel.

\section{Three neutrinos}

Oscillations of three neutrinos $(\nu_1,\nu_2,\nu_3)$, under the hypothesis of
one mass scale dominance for atmospheric $\nu$'s (equivalent to set
$m^2_1\simeq m^2_2$), are characterized by one squared mass difference,
\begin{equation}
m^2=m^2_3-m^2_{1,2}\ ,
\end{equation}
(the cases $m^2>0$ and $m^2<0$ being physically different \cite{re08}), and by
two mixing angles ($\psi=\theta_{23}\in [0,\pi/2]$ and $\phi=\theta_{13}\in
[0,\pi/2]$), describing the flavor content of the state $\nu_3$,
\begin{eqnarray}
|\langle \nu_3|\nu_e\rangle|&=&s_\phi\ ,\\
|\langle \nu_3|\nu_\mu\rangle|&=&c_\phi s_\psi\ ,\\
|\langle \nu_3|\nu_\tau\rangle|&=&c_\phi c_\psi\ ,
\end{eqnarray}
the pure $\nu_\mu\leftrightarrow\nu_\tau$ case being recovered for  $\phi=0$
(see \cite{re02} and references therein).

In order to show the bounds in the mixing parameter space, we used in
\cite{re02} a triangular representation \cite{re09,re10}, which is basically a
linear mapping in the variables ($\sin^2\psi,\sin^2\phi$). Such a
representation has the advantage of embedding unitarity by construction, but
has the disadvantage of not showing in detail the phenomenologically
interesting case of small $\phi$. We use here the alternative representation in
terms of $(\tan^2\psi,\tan^2\phi)$ in logarithmic scale (also introduced in
\cite{re09,re10}), which expands the small $\phi$ region while preserving
octant symmetry (when applicable).

Using SK data only, and assuming $m^2>0$, we find the best fit
$(\chi^2_{\min}=38.1)$ at $(m^2,\tan^2\psi,\tan^2\phi)= (3\times 10^{-3}{\rm\
eV}^2, 0.9, 0.01)$. The slight deviation of the best fit mixing  from the pure
$\nu_\mu\leftrightarrow\nu_\tau$ maximal mixing 
[$(\tan^2\psi,\tan^2\phi)=(1,0)$], although intriguing%
\footnote{There is no reason to have {\em exactly\/}
$(\tan^2\psi,\tan^2\phi)=(1,0)$.},
is---unfortunately---not statistically significant  $(\Delta \chi^2\lesssim
1)$. This also implies that there is no significant indication for possible
matter effects related to $\nu_e$ mixing $(\tan^2\phi>0)$ in the SK data.

Figure~3 shows the $3\nu$ volume allowed at 90\% and 99\% C.L.\ ($\Delta
\chi^2=6.25$ and 11.36 for $N_{\rm DF}=3$, respectively) in the
$(m^2,\tan^2\psi,\tan^2\phi)$ parameter space, through its projections onto the
coordinate planes. The upper limit on $\tan^2\phi$ ($\lesssim 0.35$ at 90\%
C.L.) improves the one found in \cite{re02} ($\tan^2\phi\lesssim 1$ at 90\%
C.L.\ from 33 kTy SK data), showing the steady progress of SK in confirming
dominant $\nu_\mu\leftrightarrow\nu_\tau$ mixing and in constraining additional
$\nu_e$ mixing. The 90\% C.L.\ range for $m^2$ is  $(1.6$--$7.2)\times 10^{-3}$
eV$^2$. The bounds on $\tan^2\psi$ in Fig.~3 are octant-symmetric only in the
$2\nu$ limit $\tan^2\phi\to0$ (as they should), and show a slight preference
for $\psi$ in the second octant when $\tan^2\phi>0$. Correspondingly, slightly
higher values of $m^2$ are preferred. The (weak) positive correlation between
$\tan^2\phi$ and $\tan^2\psi$ or $m^2$, however, is largely suppressed by the
inclusion of CHOOZ data, as we now discuss.

As described in \cite{re06}, we can now include the CHOOZ  reactor spectral
data (14 bins minus one adjustable normalization parameter) through a $\chi^2$
statistics reproducing the bounds of the so-called ``CHOOZ analysis A''
\cite{re05}. This improvement provides more accurate bounds in the $m^2$ region
of interest for atmospheric neutrinos. Our best fit to SK+CHOOZ data
($\chi^2_{\min}=45.7$) is reached at $(m^2,\tan^2\psi,\tan^2\phi)= (3\times
10^{-3}{\rm\ eV}^2, 0.75, 0.003)$. Once again,  the small deviation of the best
fit mixing  from  $(\tan^2\psi,\tan^2\phi)=(1,0)$ is not statistically
significant ($\Delta \chi^2\lesssim 1$).

Figure~4 shows the projections of the  $(m^2,\tan^2\psi,\tan^2\phi)$ volume
allowed by SK+CHOOZ. By comparing Fig.~3 with Fig.~4, the tremendous impact of
CHOOZ on $\nu_e$ mixing bounds becomes evident  (one order of magnitude
difference in the upper bound on $\tan^2\phi$). As expected, at the small
values of $\tan^2\phi$ allowed by the fit in Fig.~4, both the octant-asymmetry
in $\psi$ and the upper limit on $m^2$ are reduced, and the 90\% C.L.\ range
for $m^2$ becomes $(1.6$--$5.3)\times 10^{-3}$ eV$^2$.

We have also repeated the fit for the case $m^2<0$ (not shown), corresponding
to a state $\nu_3$ lighter than $\nu_{1,2}$. For negative $m^2$, we get
somewhat weaker bounds on $\tan^2\phi$ ($\lesssim 0.5$ at 90\% C.L.) for the
fit to SK data only, while the fit to SK+CHOOZ data gives results almost
identical to those in Fig.~4. This fact shows that, unfortunately, current
atmospheric+reactor data are basically unable to discriminate the sign of $m^2$
in $3\nu$ scenarios, as it was the case for pre-SK and pre-CHOOZ data
\cite{re08}.

Finally, Fig.~5 shows the SK zenith distributions computed for three
representative cases at $\tan^2\phi=0.025$ (allowed at 90\% C.L.\ by SK+CHOOZ)
and for $\psi$ both maximal ($\tan^2\psi=1$) and nonmaximal ($\tan^2\psi=1/2$
and 2). Within statistical errors, the $3\nu$ zenith distributions in Fig.~5
are hardly distinguishable from the $2\nu$ one in Fig.~1 (even more so for
$m^2<0$, not shown), the differences being at most $\sim 1.5\sigma$ in a few
bins. Therefore, there is little hope to unambiguously discover $\phi\neq 0$
(i.e., $\nu_e$ mixing) from SK atmospheric data in the near future.

\section{Four neutrinos}

We consider the $4\nu$ (3 active + 1 sterile) scenario described in 
\cite{re04}, characterized by a 2+2 mass spectrum with well-separated
atmospheric and solar doublets. We also make the simplifying assumption
\cite{re04} that the atmospheric doublet $(\nu_3,\nu_4)$ is almost decoupled
from $\nu_e$, and that the solar doublet $(\nu_1,\nu_2)$ is almost decoupled
from $\nu_\mu$. The dominant mass scale for atmospheric neutrinos is then
\begin{equation}
m^2=m^2_4-m^2_3\ ,
\end{equation}
and two mixing angles ($\psi=\theta_{23}\in [0,\pi/2]$ and $\xi=\theta_{13}\in
[0,\pi/2]$) are sufficient to describe the flavor contents of the state
$\nu_4$,
\begin{eqnarray}
|\langle \nu_4|\nu_e\rangle|&\simeq & 0\ ,\\
|\langle \nu_4|\nu_\mu\rangle|&=&s_\psi\ ,\\
|\langle \nu_4|\nu_\tau\rangle|&=&c_\psi c_\xi\ ,\\
|\langle \nu_4|\nu_s\rangle|&=&c_\psi s_\xi\ ,
\end{eqnarray}
the pure $\nu_\mu\leftrightarrow\nu_\tau$  and $\nu_\mu\leftrightarrow\nu_s$ 
cases being recovered for  $s_\xi=0$ and $s_\xi=1$, respectively. In such
scenario, the cases $m^2>0$ and $m^2<0$ are not physically different (being
equivalent under octant inversion, $\psi\to\pi/2-\psi$ \cite{re04}), and we
take $m^2>0$.

Figure~6 shows the bounds obtained by a fit to SK data at 90\% and 99\% C.L.\
($N_{\rm DF}=3$) in the parameter space  $(m^2,\tan^2\psi,\tan^2\xi)$. The best
fit point $(\chi^2_{\min}=38.1)$ is reached at  $(m^2,\tan^2\psi,\tan^2\xi)=
(3\times 10^{-3}{\rm\ eV}^2, 0.76, 0.1)$ but,  once again,  the preferred
mixing differs from  $(\tan^2\psi,\tan^2\phi)=(1,0)$ by less than one unit in
$\Delta \chi^2$. Notice that in the $(m^2,\tan^2\psi)$ plane, the $4\nu$
projected bounds of Fig.~6 are very similar to the $3\nu$ ones in Fig.~4,
implying that the current limits on the ``$2\nu$'' subset of parameters
$(m^2,\tan^2\psi)$ are rather stable even by making allowance for additional
$\nu_s$ or $\nu_e$ mixing.

Concerning the $(\tan^2\psi,\tan^2\xi)$ plane in Fig.~6, the upper bounds on
$\tan^2\xi$ indicate that pure  $\nu_\mu\leftrightarrow\nu_s$ oscillations
$(\tan^2\xi\to\infty)$ are disfavored as compared with pure 
$\nu_\mu\leftrightarrow\nu_\tau$ oscillations $(\tan^2\xi\to 0)$, in agreement
with \cite{re11,re12}, although large $\nu_s$ mixing is not excluded yet. In
particular, the current upper limit from Fig.~6 ($\tan^2\xi\lesssim 4$ at 90\%
C.L.) is even slightly {\em weaker\/} than the one we found with smaller
statistics in \cite{re04} ($\tan^2\xi \lesssim 2$). The reason can be traced to
a  peculiar feature of the latest UP$\mu$ data, namely, the flatness of the
muon suppression pattern in the four UP$\mu$ bins at
$\cos\theta\in[-0.7,-0.4]$, as described in Fig.~7.

Figure~7 shows the SK zenith distributions for three representative $4\nu$
cases with sizable $\nu_s$ mixing ($\tan^2\xi=1$). As it is well known (and
evident from a comparison of Fig.~7 with Fig.~1), additional $\nu_s$ mixing for
atmospheric neutrinos tends to reduce the muon suppression and, in particular,
tends to flatten the normalized UP$\mu$ distribution. Although the SK UP$\mu$
data do prefer a mean positive slope rather than a flat suppression, the four
bins in the zenith range  $\cos\theta\in[-0.7,-0.4]$ happen to favor a locally
flat distribution. This current feature might be just a statistical fluctuation
but, at present, it plays some role in global fits, where it tends to weaken
the rejection of ``flat'' distributions (i.e., of sizable $\nu_s$ mixing), as
compared with previous UP$\mu$ data \cite{re04}.

The SK Collaboration has also presented additional (preliminary)  indication in
favor of $\nu_\mu\leftrightarrow\nu_\tau$ mixing coming from statistical
$\nu_\tau$ appearance in selected event samples \cite{re01}. It seems possible
to isolate an excess of about $100\pm 50$ $\tau$-like events, to be compared
with standard $\nu_\mu\leftrightarrow\nu_\tau$ expectations of $\sim 100$
\cite{re01}. Taken at face value, such numbers imply an additional $\sim
2\sigma$ evidence ($\Delta \chi^2 \simeq 4$) in favor of pure
$\nu_\mu\leftrightarrow\nu_\tau$  ($s^2_\xi=0$) as compared with pure
$\nu_\mu\leftrightarrow\nu_s$  ($s^2_\xi=1$).  We have then roughly
parametrized the SK ``tau appearance''  signal by adding a penalty function
$\Delta \chi^2=4 s^2_\xi$ in the $4\nu$ fit. We get an ``improved'' upper bound
$\tan^2\xi\lesssim 1.5$ at 90\% C.L.\ ($N_{\rm DF}=3$), to be compared with
$\tan^2\xi\lesssim 4$ in Fig.~6 (without penalty function). On the one hand,
this seems to indicate that there is certainly room to refine current bounds on
additional $\nu_s$ mixing in SK; on the other hand, our analysis shows that
large $\nu_s$ mixing  (e.g., a fifty-fifty admixture of $\nu_\tau$ and $\nu_s$
at $\tan^2\xi\simeq 1$) is not yet excluded at present. Therefore,
compatibility with complementary solar neutrino bounds on $\tan^2\xi$
\cite{re13} is still possible, as it was the case for previous SK data
\cite{re04}.

\section{Two, three, and four neutrino summary}

The bounds on dominant and subdominant mixing found in the previous sections
can be conveniently summarized in one single plot, as shown in Fig.~8. The left
panel shows the bounds on the $(m^2,\tan^2\psi)$ parameters for pure
$\nu_\mu\leftrightarrow\nu_\tau$ mixing (equivalent to our $3\nu$ scheme for
$\tan^2\phi=0$ or to our $4\nu$ scheme for  $\tan^2\xi=0$). As discussed
before, such bounds are not significantly altered by additional $\nu_e$ mixing
($\tan^2\phi>0$) or by additional $\nu_s$ mixing ($\tan^2\xi>0$), and thus they
hold also in the global $3\nu$ and $4\nu$ fits with good accuracy. For such
reason, the $2\nu$ bounds in the left panel are formally obtained for $N_{\rm
DF}=3$, so as to match those in the middle and right panels. The middle panel
shows the  $3\nu$ bounds on additional $\nu_e$ mixing ($\tan^2\phi>0$), with
and without CHOOZ.  Finally, the left panel shows the $4\nu$ bounds on
additional $\nu_s$ mixing ($\tan^2\xi>0$). Such synthetic figure represents the
main result of our analysis.

\section{Two mass scales}

The analyses in the previous sections  are based on the assumption that
atmospheric neutrino oscillations are driven by only one mass scale $(m^2)$.
This is not necessarily the case, especially if one takes the ``solar'' squared
mass difference in the upper range allowed by the data (see, e.g.,
\cite{re14}), provided that one accepts an averaged or quasiaveraged solar 
neutrino survival probability \cite{re06}.  Concerning atmospheric neutrinos,
two mass scales in the range $\sim 10^{-3}$ eV$^2$ were shown to provide
acceptable fits to previous SK+CHOOZ data  \cite{re15}. Here we show, through a
specific example, that such possibility is not yet excluded by the latest data.

Let us consider the specific $3\nu$ case shown in Fig.~9, characterized by two
equal squared mass differences ($\Delta m^2_{32}=\Delta m^2_{21}=0.7\times
10^{-3}$) eV$^2$, and by mixing angles%
\footnote{In standard notation, $\phi=\theta_{13}$, $\psi=\theta_{23}$,
and $\omega=\theta_{12}$.}
 $(\tan^2\phi,\tan^2\psi,\tan^2\omega)=(0,1,2)$, giving the flavor composition
of mass eigenstates shown in the same figure. The spectrum in Fig.~9 might be
called ``democratic'', since it maximally violates the usual ``hierarchical''
approximation ($\Delta m^2_{21}\ll \Delta m^2_{32}$). Notice that $\nu_e$
oscillations are driven only by $\Delta m^2_{21}$, which is purposely chosen
just below the current CHOOZ bounds \cite{re05}. The solar neutrino survival
probability is then $P(\nu_e\to\nu_e)\simeq 1-\frac{1}{2}\sin^2 2\omega \simeq
5/9\sim 1/2$, up to small (quasiaveraged \cite{re06}) corrections.

Figure~10 shows the SK zenith distributions computed for the democratic
scenario in Fig.~9, and corresponding to $\chi^2=50.9$ for the fit to SK data
only ($\chi^2=61.2$ if CHOOZ data are also included). Although such value is
significantly higher than in the best-fit $2\nu$ case of Fig.~1, it is still
acceptable from the point of view of goodness of fit. All in all, the curves in
Fig.~10 provide a globally acceptable ``fit-by-eye'',   with moderate
departures from the data along the horizontal direction ($\cos\theta\sim 0$)
for the MG$\mu$ and US$\mu$ samples. Therefore, if one accepts an almost
constant $(\sim 1/2)$ suppression as explanation of the solar neutrino deficit,
$3\nu$ scenarios with two comparable mass scales appear to represent a viable
possibility in the current atmospheric neutrino phenomenology. Needless to say,
the most general analysis of such cases (depending on all the $3\nu$
mass-mixing parameters) would be rather intricated, and is postponed to a
future work.

\section{Conclusions}

In the context of the latest (79.5 kTy) SK atmospheric $\nu$ data, we have
concisely reviewed the status of dominant  $\nu_\mu\leftrightarrow\nu_\tau$
oscillations (including the case of nonstandard dynamics, Fig.~2) and of
subdominant $\nu_e$ and $\nu_s$ mixing (in $3\nu$ and $4\nu$ schemes,
respectively). In the $3\nu$ case we have applied an improved CHOOZ analysis.
The main $2\nu$, $3\nu$, and $4\nu$ results are discussed separately and then
summarized in Fig.~8. Finally, we have shown (through a specific example)  that
current atmospheric neutrino data are also compatible with  oscillations driven
by two comparable mass scales.

\acknowledgments

This work was supported by INFN and by  the Italian MURST within the
``Astroparticle Physics'' project.


%
\newcommand{\InsertFigure}[2]{\newpage\begin{center}\mbox{%
\epsfig{bbllx=1.4truecm,bblly=1.3truecm,bburx=19.5truecm,bbury=26.5truecm,%
height=21.4truecm,figure=#1}}\end{center}\vspace*{-1.8truecm}%
\parbox[t]{\hsize}{\small\baselineskip=0.5truecm\hspace*{0.5truecm} #2}}
\InsertFigure{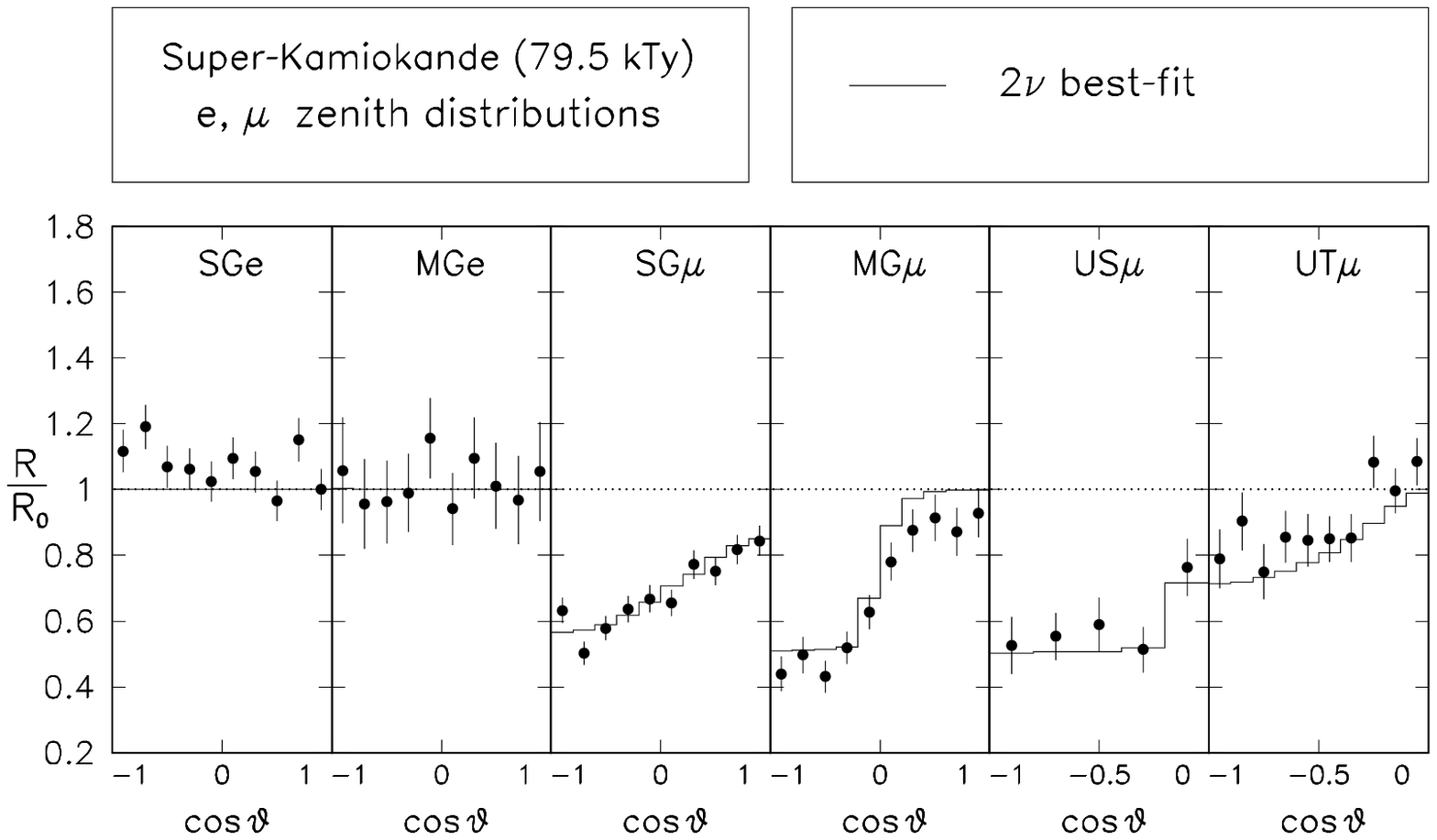}%
{Fig.~1. Super-Kamiokande zenith distributions (79.5 kTy \protect\cite{re01})
used in the analysis, normalized to no-oscillation expectations. The data set
includes sub-GeV electrons (SG$e$, 10 bins), multi-GeV electrons (MG$e$, 10
bins), sub-GeV muons (SG$\mu$, 10 bins), multi-GeV muons  (MG$\mu$, 10 bins),
upward stopping muons (US$\mu$, 5 bins), and upward through-going muons
(UT$\mu$, 10 bins), for a total of 55 data points. The error bars are
statistical only ($\pm 1\sigma$); systematic (correlated) uncertainties are
treated as in \protect\cite{re02}. The solid line is our best fit for $2\nu$
oscillations ($\chi^2=38.5$).}
\InsertFigure{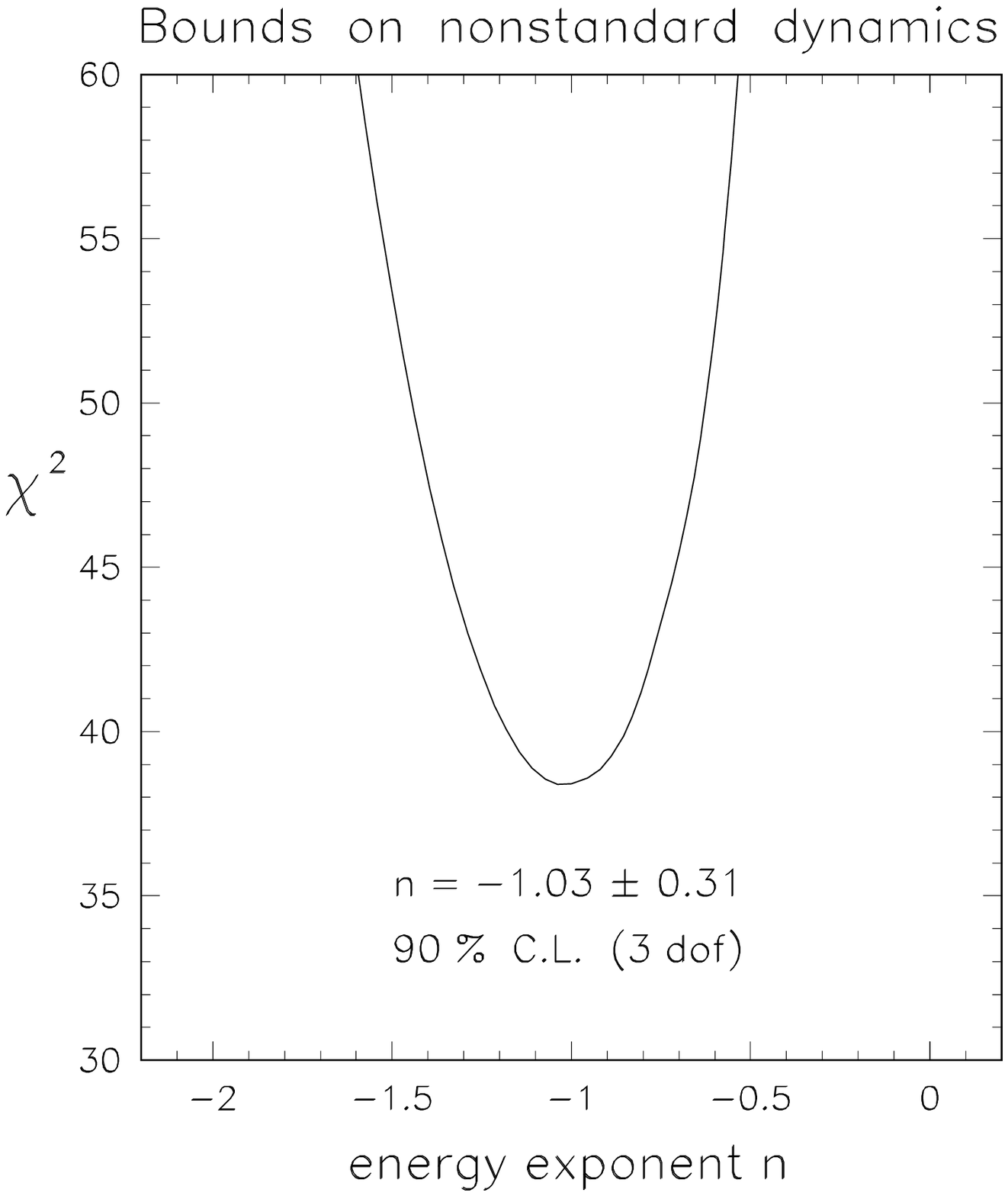}%
{Fig.~2. Dependence of the $\chi^2$ function on the neutrino energy exponent
$n$, assuming an oscillation phase proportional to $E^n$, with unconstrained
factors for the overall phase and amplitude. The only integer $n$ compatible
with the SK data is $n=-1$, corresponding to standard mass-mixing dynamics.}
\InsertFigure{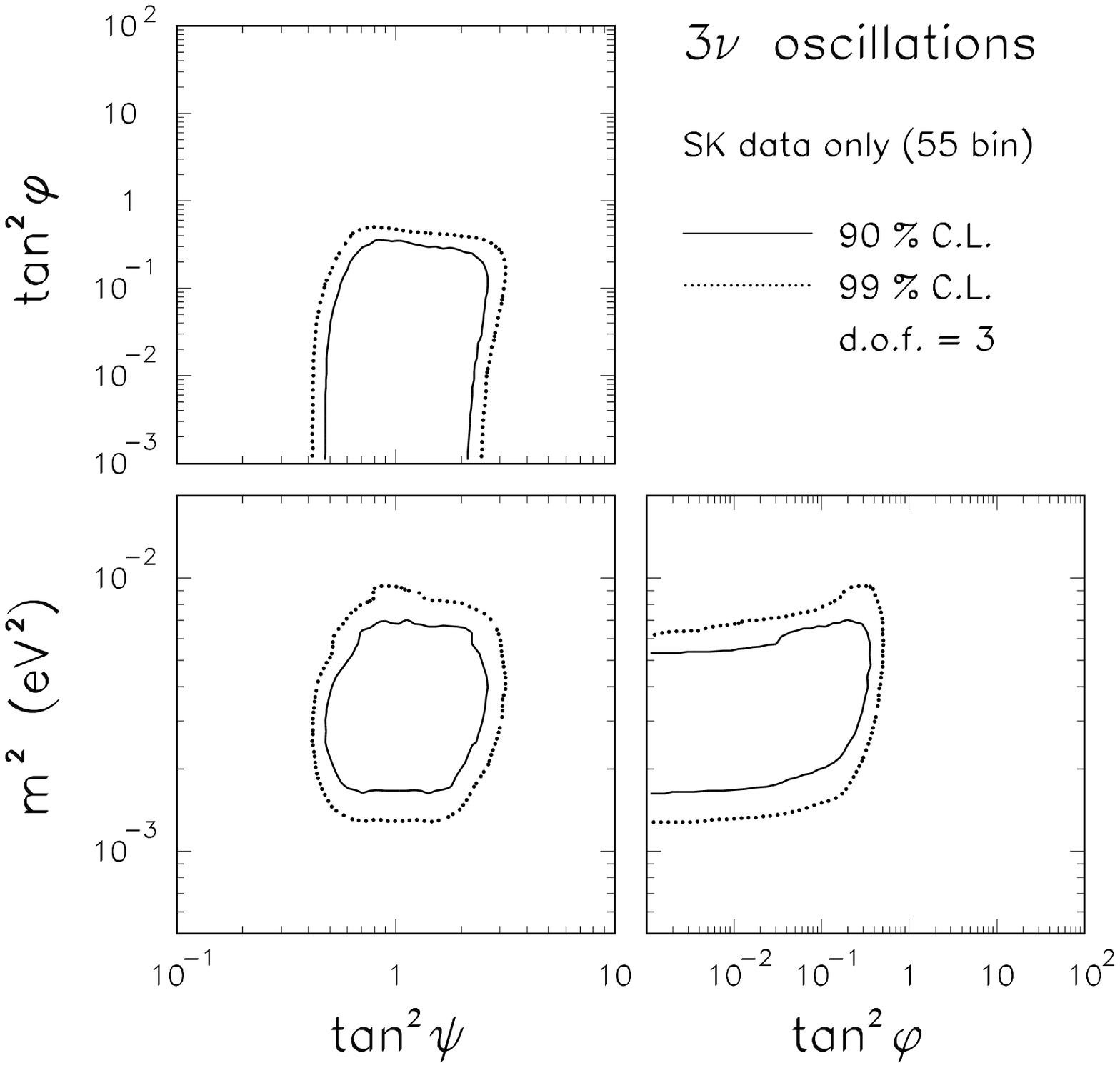}%
{Fig.~3. Projections of the regions allowed  in the $3\nu$ parameter space
$(m^2,\tan^2\psi,\tan^2\phi)$  at 90\% and 99\% C.L.\  ($\Delta \chi^2=6.25$
and 11.36 for $N_{\rm DF}=3$) onto the coordinate planes. The fit includes SK
data only (79.5 kTy). The pure $2\nu$ case of $\nu_\mu\leftrightarrow\nu_\tau$
oscillations is recovered for  $\tan^2\phi\to 0$. Nonzero values of $\phi$
parametrize $\nu_e$ mixing. }
\InsertFigure{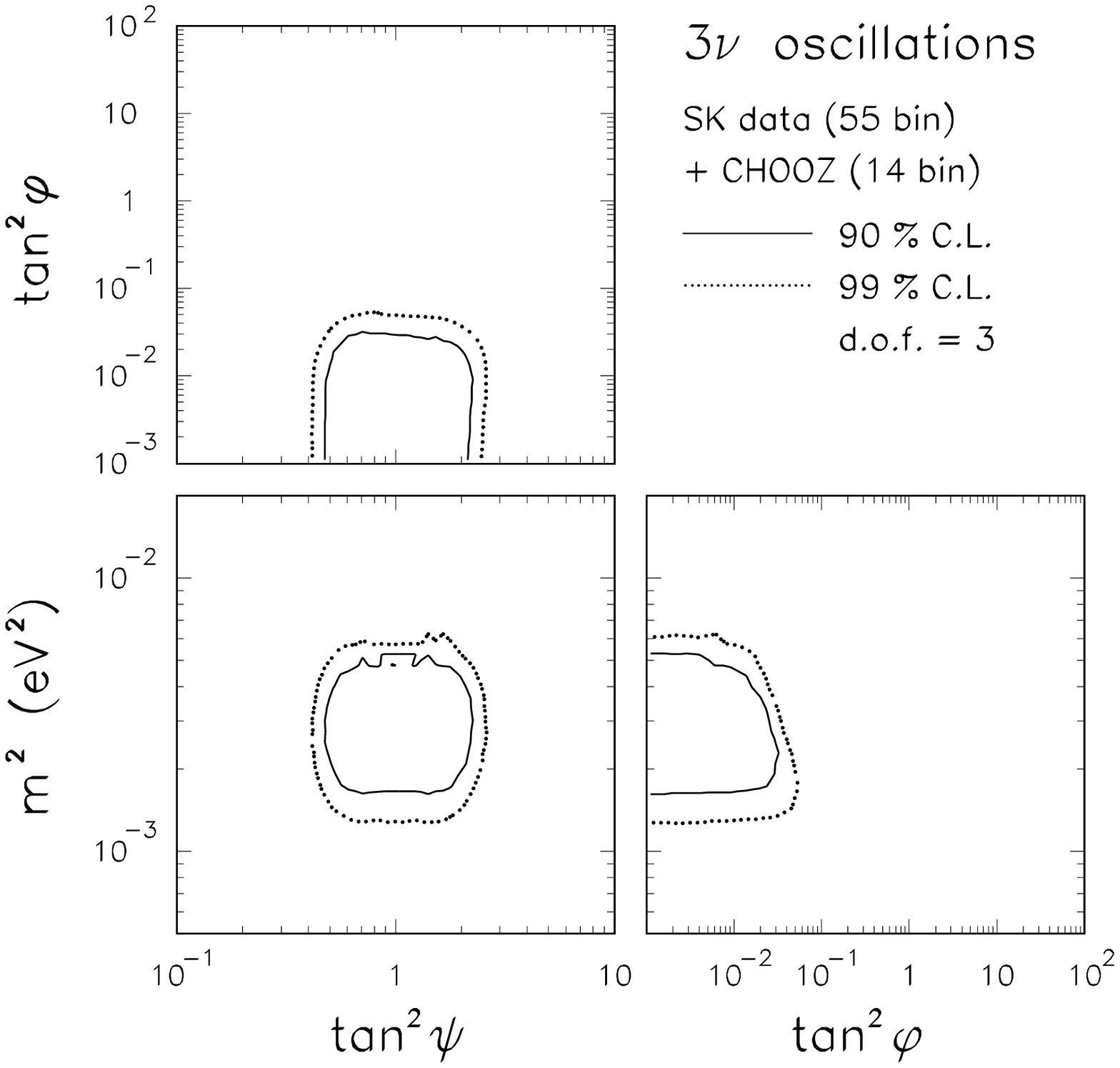}%
{Fig.~4. As in Fig.~3, but including final CHOOZ positron spectra
\protect\cite{re05} (14 data points minus one adjustable normalization
factor).}
\InsertFigure{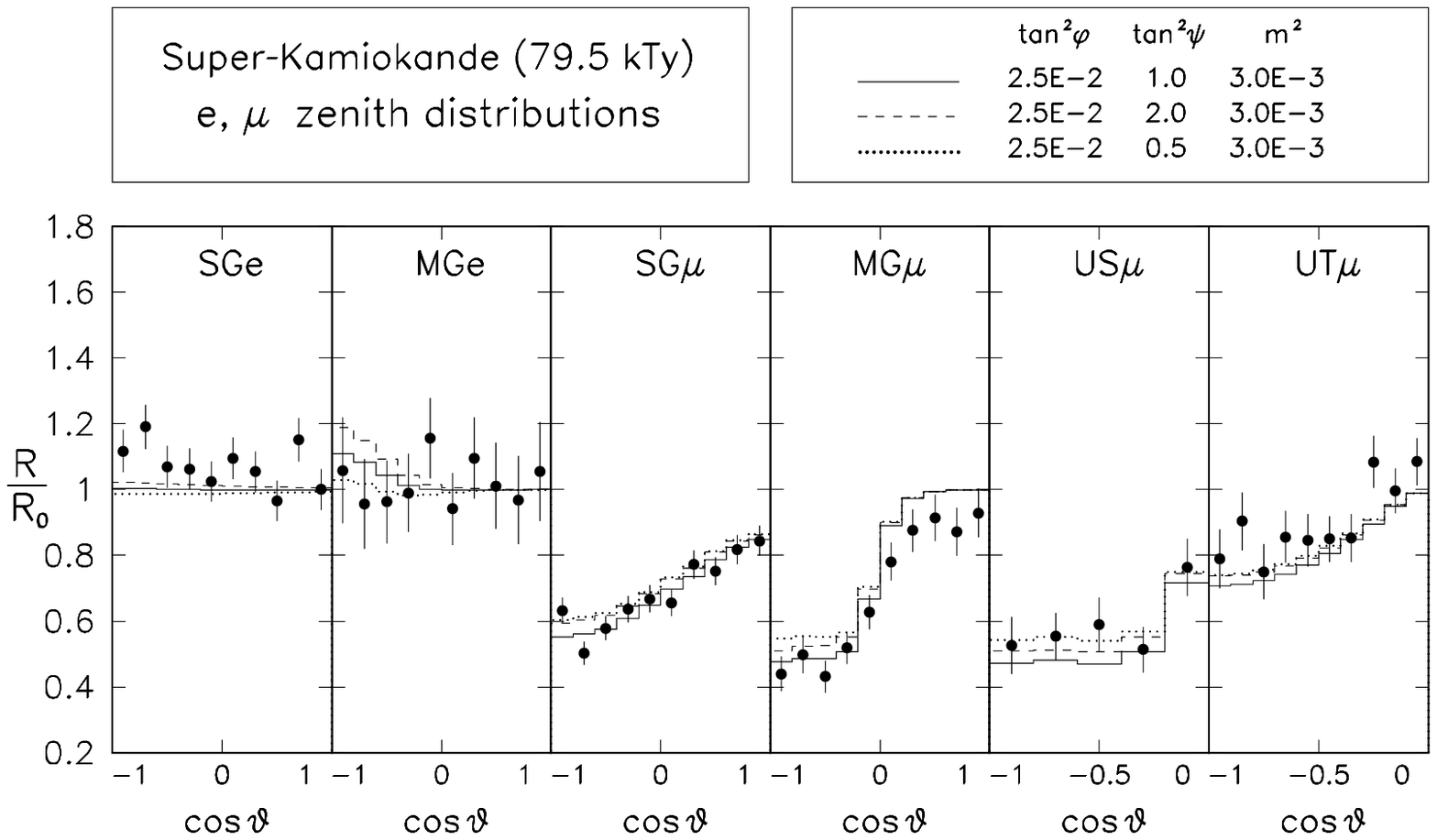}%
{Fig.~5. Zenith distributions for three representative $3\nu$ cases with 
$\tan^2\phi=2.5\times 10^{-2}$, allowed at 90\% C.L.\ by SK+CHOOZ. Notice the
distortion of the MG$e$ distribution. Such distortion would be somewhat smaller
for negative $m^2=-3\times 10^{-3}$ eV$^2$ (not shown).}
\InsertFigure{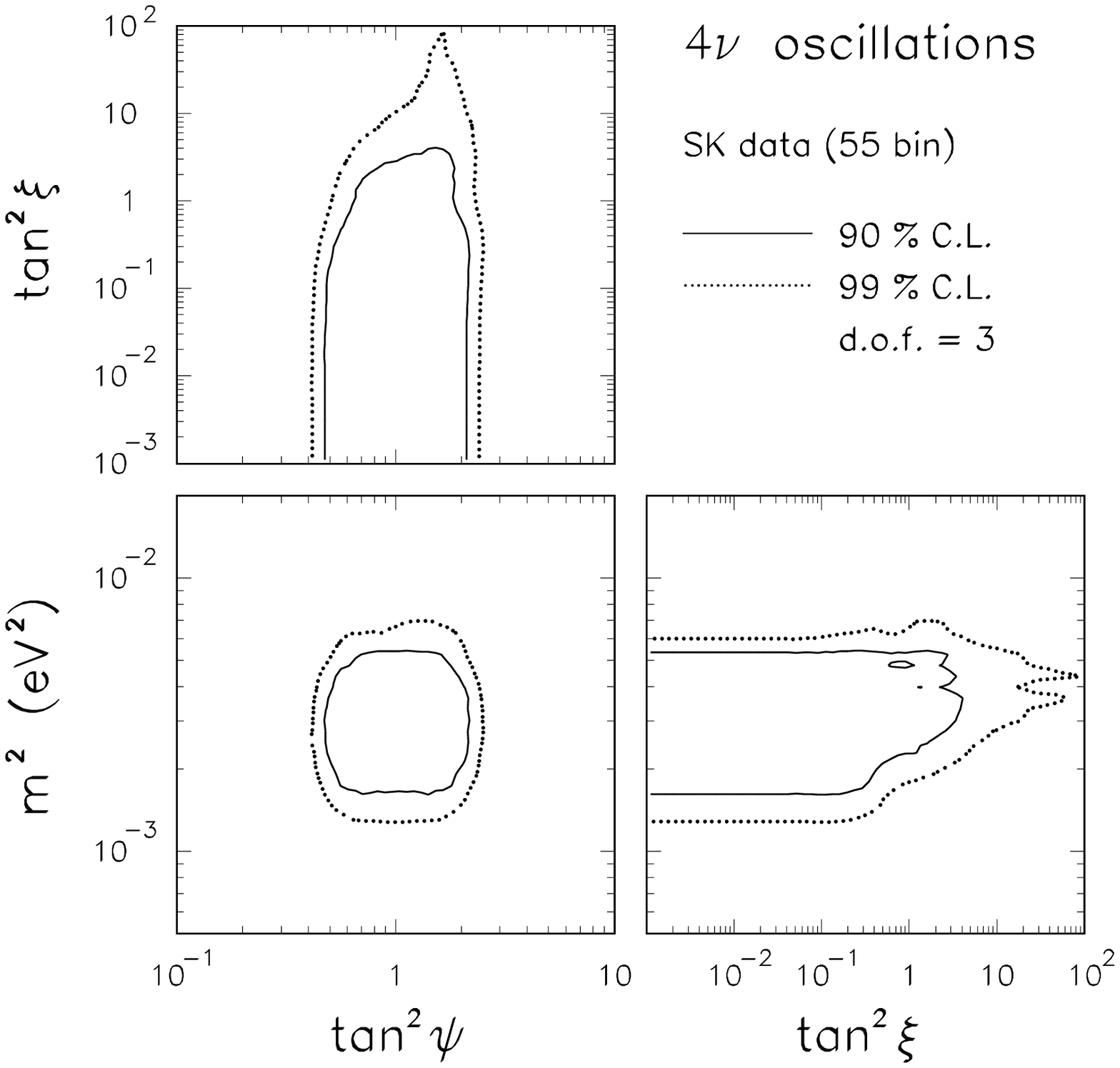}%
{Fig.~6. Projections of the regions allowed  in the $4\nu$ parameter space
$(m^2,\tan^2\psi,\tan^2\xi)$ at 90\% and 99\% C.L.\  ($\Delta \chi^2=6.25$ and
11.36 for $N_{\rm DF}=3$)  onto the coordinate planes. The fit includes SK data
only (79.5 kTy). The pure  $\nu_\mu\leftrightarrow\nu_\tau$ case 
($\tan^2\xi\to 0$) is clearly preferred over  the pure
$\nu_\mu\leftrightarrow\nu_s$ case  ($\tan^2\xi\to \infty$). However, sizable
$\nu_s$ mixing  [$\tan^2\sim O(1)$] is allowed in addition to
$\nu_\mu\leftrightarrow\nu_\tau$ oscillations.}
\InsertFigure{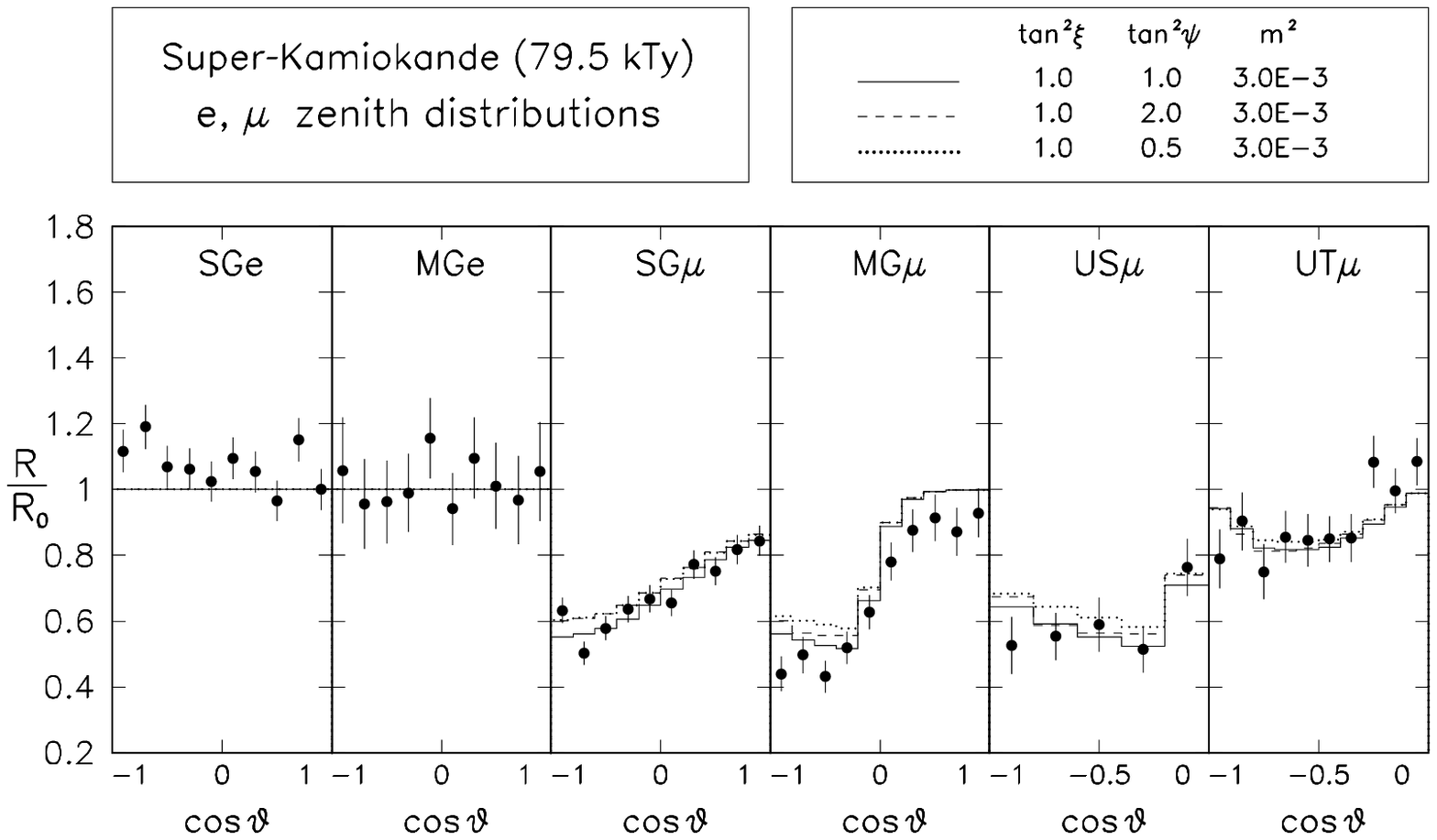}%
{Fig.~7. Zenith distributions for three representative $4\nu$ cases with 
sizable $\nu_s$ mixing ($\tan^2\xi=1$), allowed at 90\% C.L.\ by SK data.
Notice the reduced suppression in the muon samples, as compared with  pure
$\nu_\mu\leftrightarrow\nu_\tau$ mixing in Fig.~1.}
\InsertFigure{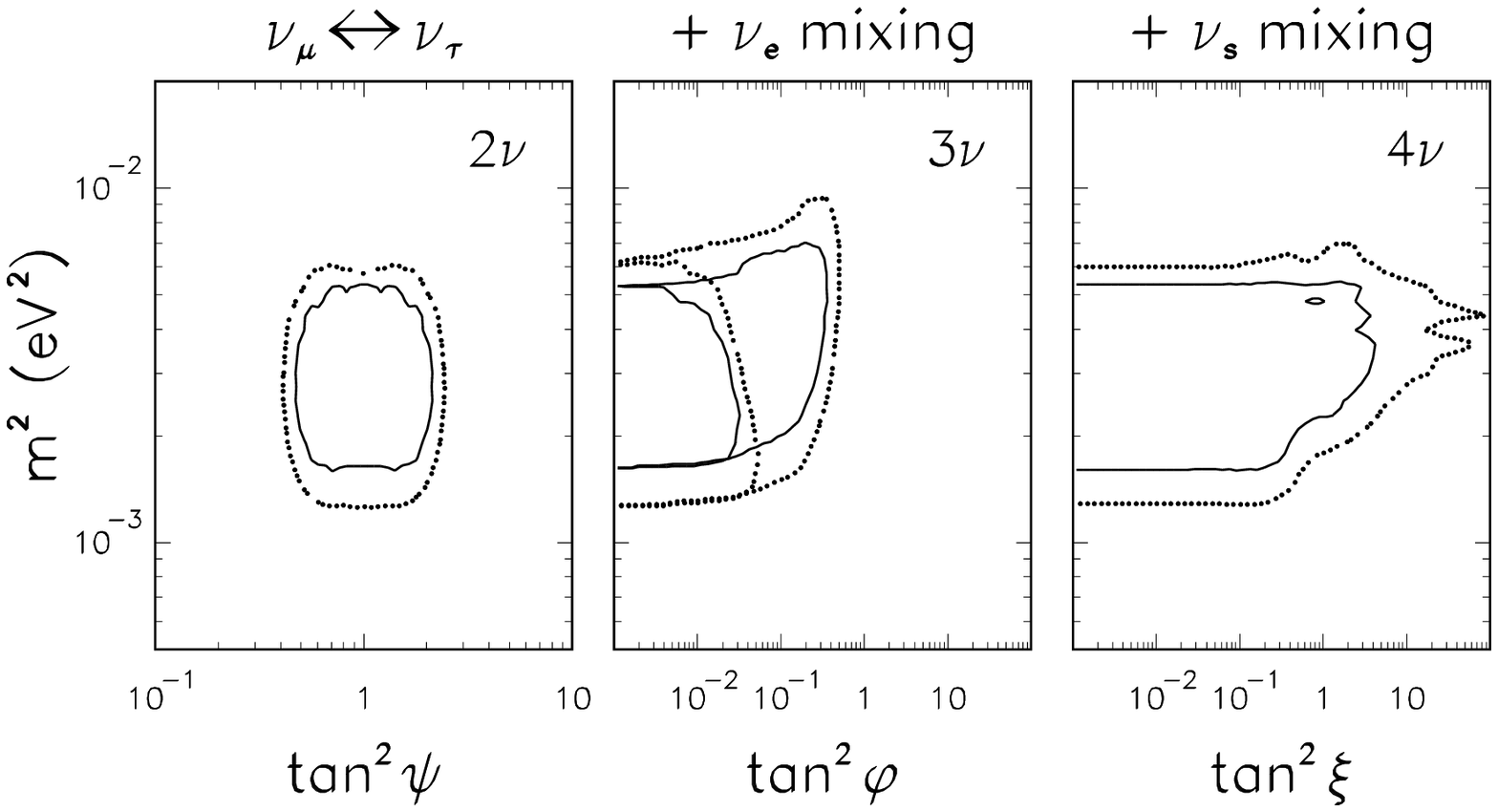}%
{Fig.~8. Summary of $2\nu$, $3\nu$, and $4\nu$ bounds at 90\% and 99\% C.L.\ on
the mass-mixing parameters from SK data. Left panel: Bounds on
$(m^2,\tan^2\psi)$ for pure $\nu_\mu\leftrightarrow\nu_\tau$ mixing (i.e., for
$\tan^2\phi=0=\tan^2\xi$). Middle panel: Bounds on additional $\nu_e$ mixing
(parametrized by $\tan^2\phi>0$) in $3\nu$ scenarios, both without CHOOZ (see
also Fig.~3) and with CHOOZ (see also Fig.~4). Right panel: Bounds on
additional $\nu_s$ mixing  (parametrized by $\tan^2\xi>0$) in $4\nu$ scenarios
(see also Fig.~6). All the bounds are derived for $N_{\rm DF}=3$, including
those in the left panel.}
\InsertFigure{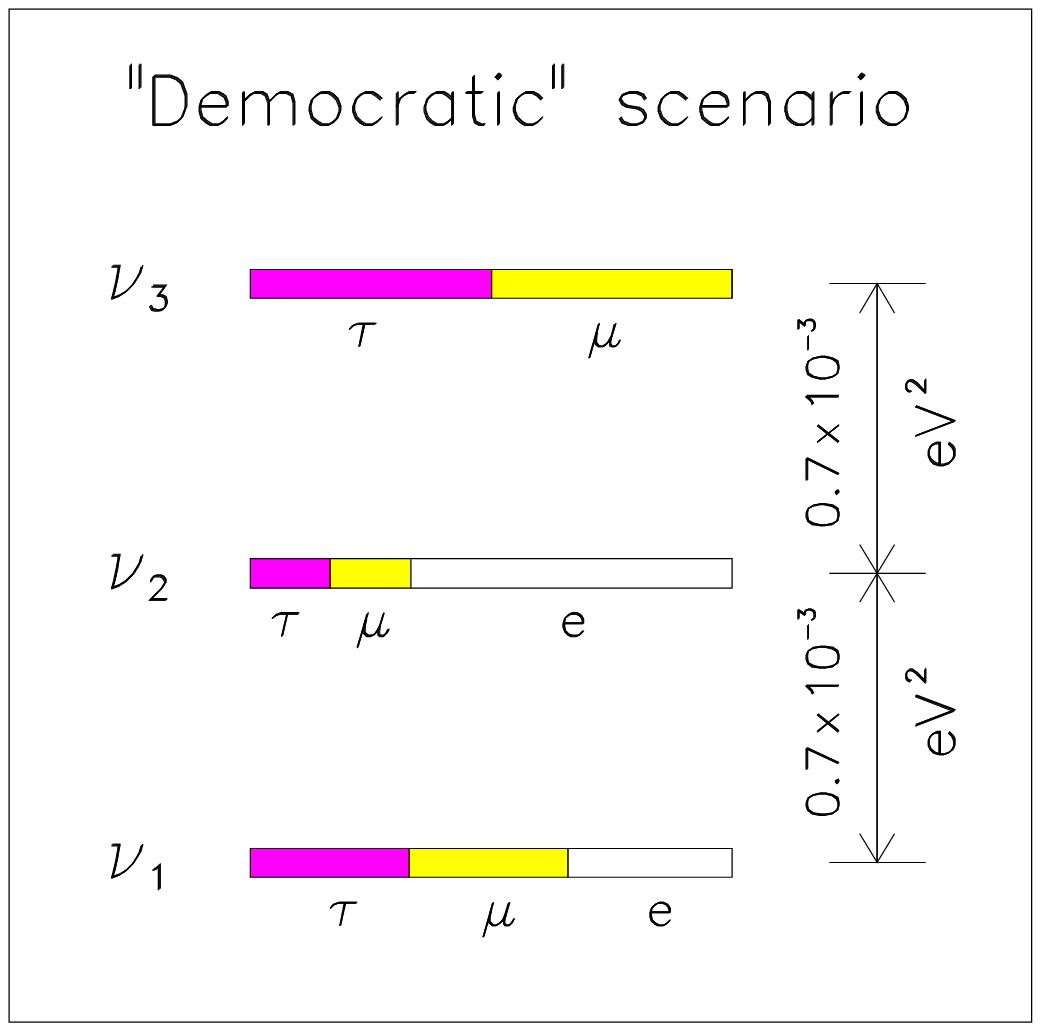}%
{Fig.~9. A ``democratic'' (i.e., nonhierarchical) $3\nu$ scenario with two
equal squared mass differences ($\Delta m^2_{32}=\Delta m^2_{21}= 0.7\times
10^{-3}$ eV$^2$) and with $(\nu_e,\nu_\mu,\nu_\tau)$ flavor content as follows:
$(1/3,1/3,1/3)$ for $\nu_1$, $(1/6,1/6,1/6)$ for $\nu_2$, and $(0,1/2,1/2)$ for
$\nu_3$. Notice that $\nu_e$ disappearance is driven only by $\Delta m^2_{21}$
(just below the CHOOZ sensitivity \protect\cite{re05}), and gives
$P(\nu_e\to\nu_e)\simeq 5/9$ for solar neutrinos (up to small quasiaveraged
oscillation corrections).}
\InsertFigure{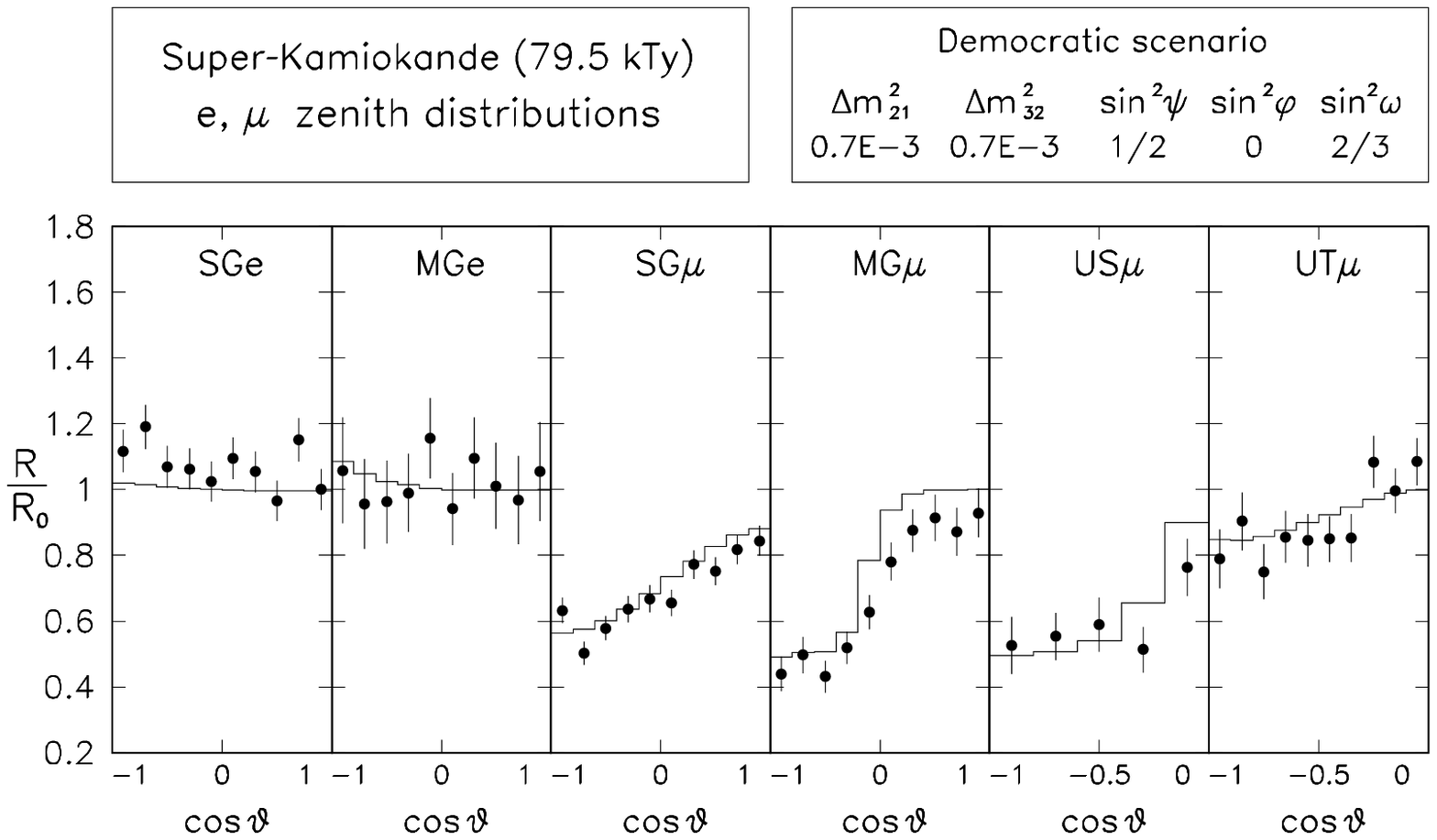}%
{Fig.~10. Zenith distributions for the democratic scenario in Fig.~9, giving
$\chi^2=50.9$ for the fit to SK data ($\chi^2=61.2$ for the fit to SK+CHOOZ).}

\eject
\end{document}